\documentclass[aps,pra,twocolumn,amsmath,showpacs,amssymb,superscriptaddress]{revtex4-1}
\usepackage{graphicx}  
\usepackage{dcolumn}   
\usepackage{bm}        
\usepackage{amssymb}   

\newcommand{\be}{\begin{equation}}
\newcommand{\ee}{\end{equation}}
\newcommand{\bea}{\begin{eqnarray}}
\newcommand{\eea}{\end{eqnarray}}

\usepackage{setspace}
\usepackage{amsmath}
\usepackage{amssymb}
\usepackage{fancybox}
\usepackage{listings}
\usepackage{url}

\begin{document}

\title{Molecular spectra in collective Dicke states}
\author{Eran Sela}
\affiliation{Raymond and Beverly Sackler School of Physics and Astronomy, Tel-Aviv University, Tel Aviv, 6997801, Israel}
\author{Victor Fleurov}
\affiliation{Raymond and Beverly Sackler School of Physics and Astronomy, Tel-Aviv University, Tel Aviv, 6997801, Israel}
\author{Vladimir A. Yurovsky}
\affiliation{School of Chemistry, Tel Aviv University, Tel Aviv, 6997801, Israel }

\begin{abstract}
We study a model describing competition of interactions between $N$ two-level systems (TLSs)
against decoherence. We apply it to analyze dye molecules in an optical microcavity, where molecular vibrations provide a local source for decoherence. Most interesting is the case when decoherence strongly affects each individual TLS, \emph{e.g.} via broadening of emission lines as well as vibrational satellites, however its influence is strongly suppressed for large $N$ due to the interactions between TLSs. In this interaction dominated regime we find unique signatures in the emission spectrum, including strong $\mathcal{O}(\sqrt{N})$ level shifts, as well as $1/N$ suppression of both the decoherence width and of the vibrational satellites.
\end{abstract}

\pacs{42.50.Pq, 03.75.Hh, 67.85.Hj}

\maketitle
\section{Introduction}
Dynamics of quantum two-level systems (TLS) has always been at the focus of interest,
but recently it has attracted increased attention because of ideas of quantum computing~\cite{QC}. A crucial requirement is the
preservation of phase coherence in the presence of a noisy \emph{environment}. The resulting spin-boson
models have been extensively studied (see the reviews~\cite{Leggett87,Weiss99}). In many systems the \emph{interactions} between TLSs lead to collective behavior, such as Dicke superradiance~\cite{Dicke}.

\begin{figure}
\centering
\includegraphics*[width=1.2\columnwidth]{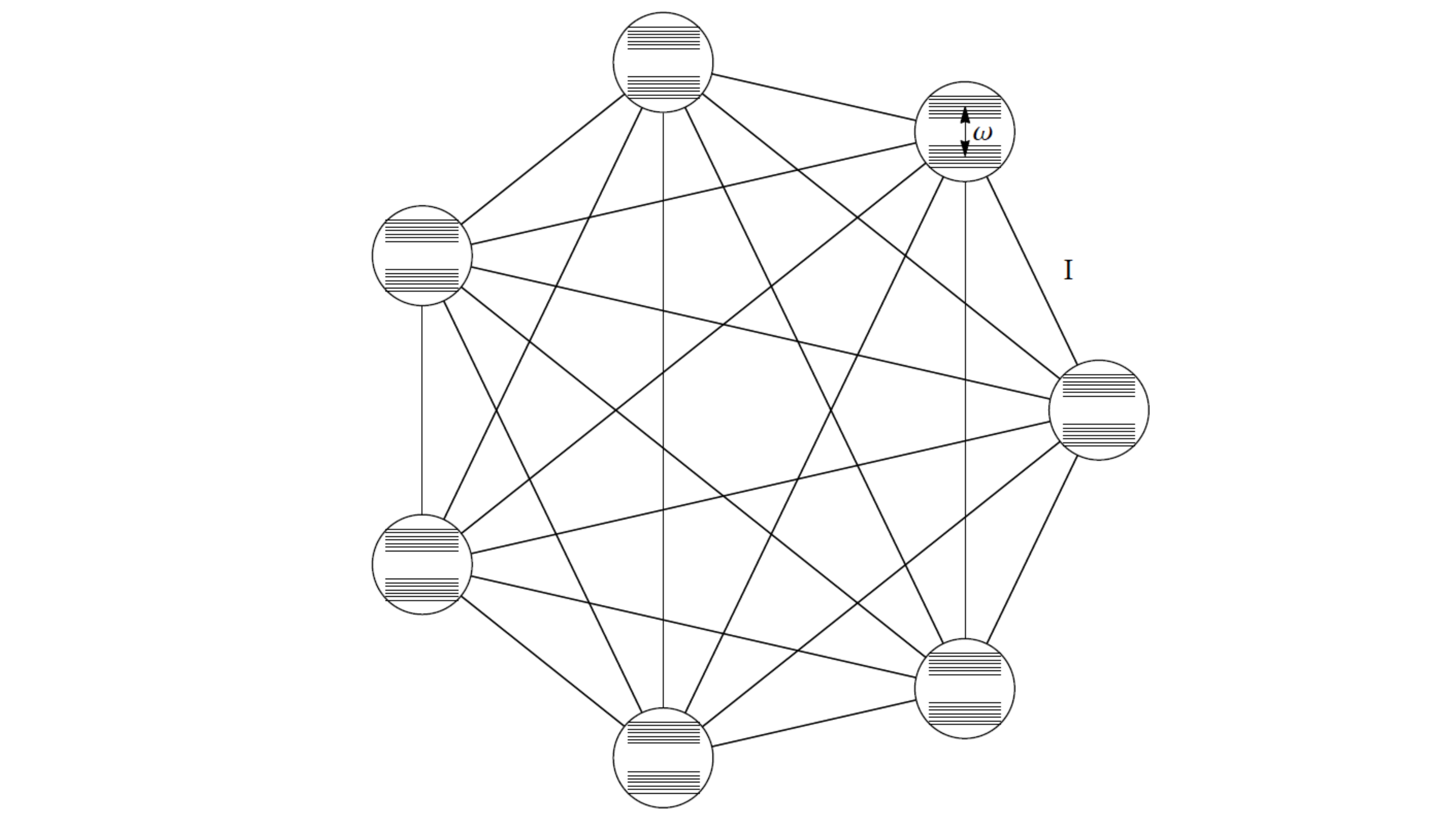}
\caption{
(Color online) Maximally connected graph describing $N$ TLSs, with all-to-all interaction $I$, where each TLS is superimposed on a quasicontinuous set of vibration levels representing local environments and internal vibrational-rotational degrees of freedom of the molecules.   }
\label{Lipkinmodel}
\end{figure}

In this article we describe the competition between local decoherence on one hand and interaction between $N$ TLSs, see Fig.~\ref{Lipkinmodel}, on the other. We consider molecules in a two-dimensional optical microcavity. A TLS excitation may hop between molecules via emission and absorption of virtual cavity photons. These are effectively two-dimensional  massive bosons~\cite{Iacopo}, and as a result the interaction acquires a finite length scale~$\ell$~~\cite{Sela14}. In this paper we assume that there is a large number $N$ of molecules within this interaction range.

As a starting point we observe that as a result of interactions, the ground state becomes a large quantum superposition of states with a given total number of excited TLSs $N_{ex} \le  N$ coherently shared between the $N$ TLSs, referred to as ``Dicke state". Under the simplifying assumption of a constant all-with-all interaction $I$, see Fig.~\ref{Lipkinmodel}, our model reduces to the (isotropic version of the) Lipkin-Meshkov-Glick (LMG) model~\cite{LMG}. We then can use language of spin-states to exploit the resulting approximate permutation symmetry, where the collective Dicke state corresponds to the ``large spin" state.

The energy difference between these many-body states depends on an interaction parameter, $I$, and also scales with $N$. Such interactions between a small number ($N=2,3$) of qubits have been implemented \emph{e.g.} in superconducting circuits~\cite{Majer2007,Fink2009} and it was suggested that these systems could in principle be scalable to larger $N$ and realize the LMG model~\cite{Larson10}. The question then is the fate of these states in the presence of a noisy environment.

Our model is relevant for the dye-filled microcavity experiments of Klaers \emph{et. al.}~\cite{Klaers10}.
Each dye molecule contains both an electronic excitation approximated by a TLS, \emph{and} a set of vibrational states, see Fig.~\ref{Lipkinmodel}. Besides, the molecules are coupled to the environment--- phonons of the solvent or substrate.

As we calculate, the coupling to the environment allows for transitions between the various collective states (\emph{e.g.} between the largest spin state and smaller spin states) or equivalently to a randomization of the phases in the quantum superposition state. Yet, despite the local decoherence which could be so strong~\cite{Martini00} as to prevent coherent behavior of a single TLS, it is reasonable to think that since each TLS is coupled  to $\mathcal{O}(N)$ other molecules, for large enough $N$ decoherence will become subdominant perturbation compared to the many-body interaction~\cite{Sela14}. In this paper we substantiate this idea with explicit calculations of the collective-level decay rates and their consequences for the molecular spectra. Conventionally, molecular spectra contain broadening and satellite vibrational peaks with associated Franck-Condon effect.
We find that when a collective state forms, the emission lines shape has (i) a $1/N$ suppression of the width, (ii) a $1/N$ suppression of vibronic satellites, and (iii) $\mathcal{O}(\sqrt{N})$ shifts of the position of the peak. These effects directly imply that transitions observed in the emission are not intra- but rather inter-molecular processes.

The essence of the competition between decoherence and interaction can be understood as follows: for a single TLS the effect of decoherence is a fluctuating phase $\phi$ in the quantum superposition state $|\uparrow \rangle + e^{i \phi} |\downarrow \rangle$ where $|\uparrow \rangle, |\downarrow \rangle$ are ground- and excited states of a single TLS. This phase does not change the energy. However, once multiple TLSs interact, the phase does modify the energy: for example for $N=2$ TLSs, one can consider a quantum superposition $|\uparrow \downarrow \rangle + e^{i \phi} |\downarrow \uparrow  \rangle$; in this case $e^{i \phi} = +1$ and $-1$ correspond to triplet and singlet spin states, respectively, which have different interaction energies. For sufficiently large $N$ such energy differences scale as $N$ and can exceed the thermal energy that can be supplied by the environment, and hence these ``large spin states" become stable against decoherence.

Many-body states with defined total spins can appear also in gases of ultracold spinor atoms, with no coupling to the  cavity modes. A mechanism of their stabilization, based on quantum interference, is proposed in Ref.~\onlinecite{yurovsky16}.

The paper is organized as follows. In Sec.~\ref{se:model} we present the model, its collective states, and its relation to the familiar Dicke model. Then, in Sec.~\ref{se:environment} we treat the effect of decoherence due to the environment, and compute the decay rate and associated level widths of the collective states. We show that decay to other spin states becomes negligible for large enough effective interaction parameter $(N \cdot I)$ exceeding the thermal energy $k_B T$. In Sec.~\ref{se:PhysicalConsequence} we discuss consequences of the interactions in the emission spectrum, which, as we claim in Sec.~\ref{se:obs}, can be observed in experiment.
We conclude in Sec.~\ref{se:conclusion}.

\section{Model}
\subsection{Cavity photon mediated dipole-dipole interaction}
We start with a motivation of the all-with-all interaction $I$ in our central model [Eq.~(\ref{eq:Hmodel})]. Consider a two-dimensional optical cavity created between two mirrors of area $L^2$ and separation $d$ (\emph{c.f.} Fig.~1 of Ref.~\onlinecite{Sela14}). The free space dispersion relation $\nu_k = c\sqrt{k_x^2+k_y^2+k_z^2}$ now becomes~\cite{Iacopo}
\be
\label{eq:dispersionm}
\nu_k = \epsilon_g+\frac{c^2 k^2}{2 \epsilon_g} = \epsilon_g + \frac{k^2}{2 m} ,~~~k=(k_x,k_y),
\ee
where $\epsilon_g = c \frac{n_z \pi}{d}$ is the cutoff frequency of the cavity ($\hbar = 1$) and $n_z$ is a fixed standing wave number. Now we place molecules acting as TLSs in the cavity at positions $(x_i,y_i,z_i)=(r_i,z_i)$, $-L/2 \le x_i,y_i \le L/2$, $0 \le z_i \le d$. Our model is
\bea
\label{eq:Hcavity}
H_{cav}=\frac{\omega}{2} \sum_i \sigma_i^z+\sum_k \nu_k a^\dagger_k a_k + \sum_{k ,i}( \gamma_{ki } a_k \sigma_i^+ + h.c.),
\eea
where $a^\dagger_k$ creates a photon at mode $k$ and
\be
\label{eq:gammaki}
\gamma_{k,i}=\gamma e^{i k \cdot r_i} \sin(\pi n_z z_i/d).
\ee
We define the detuning
\be
\Delta = \epsilon_g - \omega,
\ee
and assume it to be sufficiently large and positive such that photons become virtual excitations in the cavity.

Similar to Ref.~\cite{dipoledipole}, which studied dipole-dipole interaction induced in a one-dimensional optical cavity (see also Ref.~\cite{Zeeb15}), here we consider the two-dimensional case. Consider an initial state with no photons and one excited TLS at some molecule $i_0$. The state after time $t$ is
\bea
|\psi(t) \rangle = \sum_i b_i(t) | \downarrow \downarrow \uparrow_{i} \downarrow \downarrow \rangle +  \sum_k b_k(t) |k \rangle,
\eea
where $b_i(0)=\delta_{i,i_0}$, $b_k(0)=0$, and $|k \rangle=a^\dagger_k |0 \rangle$. These amplitudes evolve according to the Schr\"odinger equation
\bea
i \dot{b}_i = \frac{\omega}{2}(2-N) b_i+\sum_k b_k \gamma_{k , i}, \nonumber \\
i \dot{b}_k = (\nu_k + \frac{\omega}{2}(-N) )b_k+\sum_i \gamma_{k,i}^* b_i.
\eea
Using $b_i(t) = b_i(E) e^{i \frac{N \omega}{2} t - i E t}$, $b_k(t) = b_k(E) e^{i \frac{N \omega}{2} t - i E t}$, solving the Schr\"odinger equation for $b_k(E)$ and substituting into the equation for $b_i(E)$, we obtain
\be
E b_i = \omega b_i - \sum_{j}I_{ij}(E)b_j,~~~I_{ij}(E)=\sum_k \frac{\gamma_{ki} \gamma^*_{kj}}{\nu_k-E}.
\ee
Using Eq.~(\ref{eq:gammaki}), we obtain after the angular integration over $k$
\bea
I_{ij}(E)=\sin(\pi n_z z_i/d) \sin(\pi n_z z_j/d) \frac{L^2 \gamma^2}{2\pi} \int^{1/d} dk k J_0(k r), \nonumber
\eea
where $r = |r_i-r_j|$ is the distance between molecules $i$ and $j$ projected to the $xy$ plane. The Bessel function $J_0$ dictates the relevant $k$-vectors to be $k<1/r$. Since the quadratic dispersion Eq.~(\ref{eq:dispersionm}) applies only for $k \ll 1/d$, the interaction between two molecules at short distances $r \ll d$ is\emph{ not} well described by our approximation. This corresponds to the 3D free space near-field interaction. For $r \gg d$ we have
\bea
\label{eq:IK}
I_{ij}(E) &=& \sin(\pi n_z z_i/d) \sin(\pi n_z z_j/d) \frac{L^2 \gamma^2 m}{\pi}K_0(r/\ell(E)), \nonumber \\
&&~~~\frac{1}{2m \ell^2(E)} = \Delta -E.
\eea
We note few points: (i) From the asymptotic behaviour of the Bessel $K$ function $K_0(x) \to \sqrt{\frac{\pi}{2 x}} e^{-x}$ at large $x$, we see that the interaction decays exponentially over the length $\ell(E)$. (ii) In general one should solve the transcendental Schr\"odinger equation for $E$. However for large detuning $\Delta \gg |E|$ one can ignore the energy dependence of $I_{ij}(E)$ and of $\ell(E)$. Hence
        \be
\label{eq:ell}
\ell = \frac{1}{\sqrt{2m \Delta}}.
\ee (iii) As long as the separation between molecules $\delta \ll d , \ell$, and if $\ell \gg d$, then typical distances between molecules exceed $d$ and we may disregard near field interaction.  (iv) The prefactors $\sin(\pi n_z z_i/d) \sin(\pi n_z z_j/d)$, which depend on the $z$-positions of the molecules will lead to randomness in the interaction $I_{ij}$. In this paper we will focus on the average effect and ignore these factors.

Under these assumptions, the starting point for this paper is the situation~\cite{Sela14} where a number of molecules $N$ whose separation is smaller than $\ell$, interact all-with-all with a nearly constant interaction $I \sim L^2 \gamma^2 m$. Due to the dimensionality mismatch between the molecules, which are spread in the three dimensional space of the cavity, as opposed to the photons that mediate the interaction, which are two dimensional due to fixed $n_z$, the number of molecules in this correlation length becomes~\cite{Sela14}
\be
\label{eq:N}
N \sim \frac{\ell^2 d}{\delta^3}.
\ee
This paper deals with effects originating of large values of this number.

\subsection{The model}
\label{se:model}
With the above motivation, we consider the model Hamiltonian $H=H_0 + V$, with $H_0= H_S + H_v$, and
\bea
\label{eq:Hmodel}
H_{S} &=& \frac{\omega}{2} \sum_{i=1}^N \sigma_i^z  - I \sum_{i, j=1}^N  \sigma_i^+ \sigma_j^- , \\ \nonumber
H_{v} &=& \sum_{i=1}^N  \sum_l E_l v_{l,i}^\dagger v_{l,i}, \\ \nonumber
  V &=& \sum_{i=1}^N  \sum_l \sigma^z_i (C_l  v_{l,i}+h.c.).\nonumber
\eea
Here $H_{S}$ describes $N$ TLSs with Pauli operators $\sigma^a_i$, $a=z,\pm$ ($i=1,...,N$). The TLSs are connected via the all-to-all coupling term $I$, which leads to collective eigenstates described below. The second term $H_{v}$ accounts for the local baths of bosonic modes (vibrations) with energies $\{E_l \}$, created by $v^\dagger_{l,i}$, with mode $l$ at TLS $i$. The bosonic modes can represent either internal vibrational-rotational molecular degrees of freedom or phonons in solvent or substrate.
We emphasize that there is one independent bath attached to each TLS and correlations between phonons interacting with different molecules are neglected. Finally, $V$ describes the (linear) coupling between each TLS and its environment, which is a $N$-spin extension of the usual spin-boson
models~\cite{Leggett87,Weiss99}.

\begin{figure}
\centering
\includegraphics*[width=1.05\columnwidth]{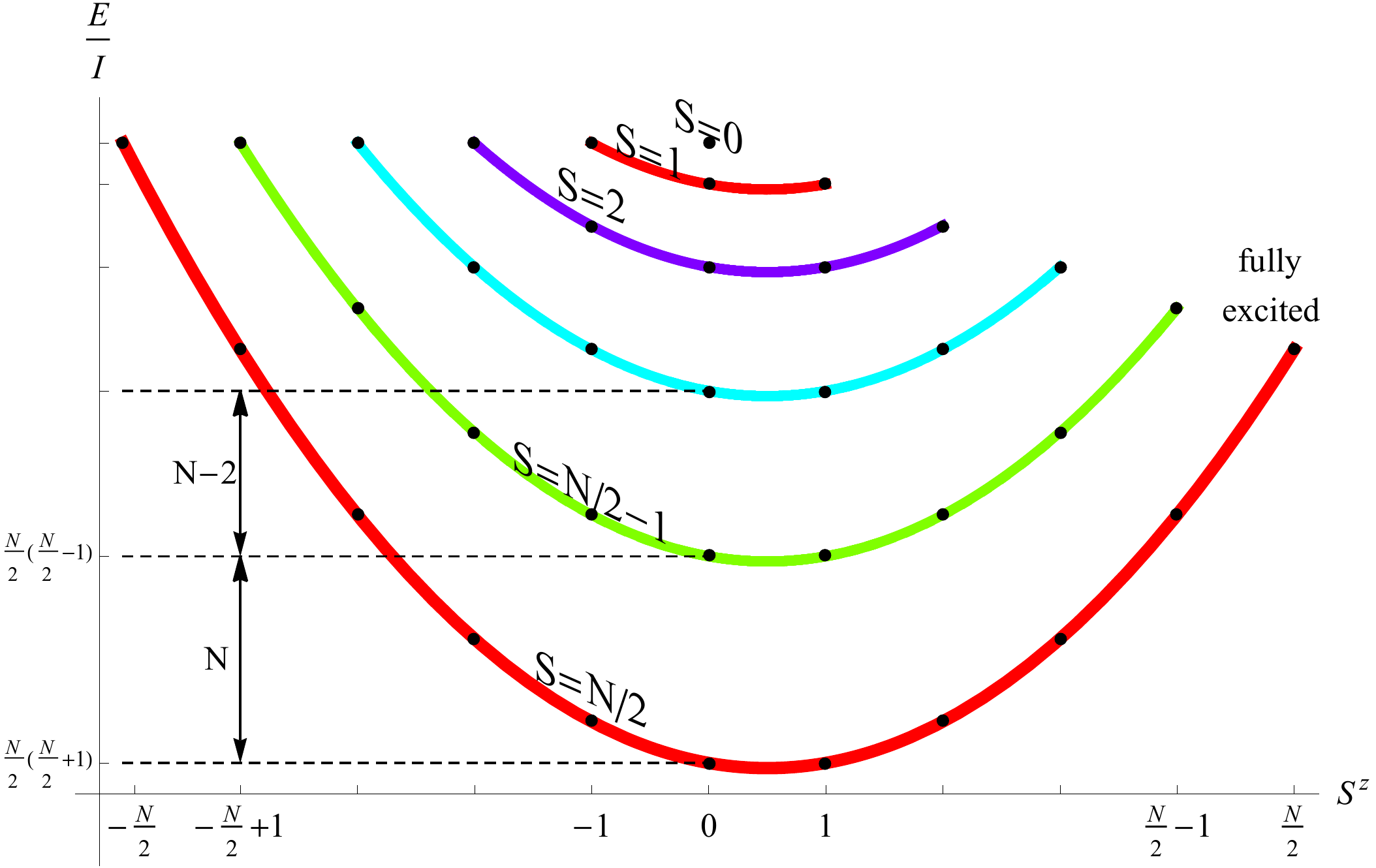}
\caption{(Color online) Energy levels of the LMG model (\ref{Lipkin}) (here $\omega = 0$). Different parabolas correspond to different total spin $S$. The largest spin state $S=N/2$ is the lowest in energy, with a gap $\delta E = N I$ to the next spin $S = N/2 - 1$ state. For nearly $50 \%$ polarization with $|S^z| < N^{1/2}$, we have an energy window $I N$, in which the large spin states are well separated (in energy) from all other states.}
\label{Lipkinlevels}
\end{figure}

Due to the permutation symmetry of the model the TLS part of the Hamiltonian $H_S$ can be conveniently written in terms of total spin operators $S^z = \frac{1}{2} \sum_{i=1}^N  \sigma^z_i$ , $S^\pm = \sum_{i=1}^N  \sigma^\pm_i$, using the relation $S^+ S^- = {\bf{S}^2} - (S^z)^2 + S^z$, as
\be
\label{Lipkin}
H_{S} = \omega  S^z - I \left( {\bf{S}}^2 - {S^z}^2 +S^z \right) .
\ee
The model (\ref{Lipkin}) when restricted to a specific total spin $S$, with ${\bf{S}}^2=S(S+1)$, is known as (a special case of) the Lipkin-Meshkov-Glick (LMG) model. However, in our system there are many different values of $S$ that $N$ TLSs can form, leading to the spectrum in Fig.~\ref{Lipkinlevels}. As we will discuss below, the environment can cause \emph{transitions }between these states.

The energy of the eigenstates of $H_S$
\be
\label{ESSz}
E_{S ,S^z}=\omega  S^z + I \left({S^z}^2 -S^z- S(S+1) \right),
\ee
depends on the total spin $S$, and on the polarization $S^z$.
The maximal total spin formed out of $N$ TLSs, each of which behaves as an elementary spin$-\frac{1}{2}$, is $S_{max}= N/2$. As can be seen in Fig.~\ref{Lipkinlevels} this large spin state gains the maximal interaction energy of $-I (N/2)(N/2+1)$ when $I > 0$. The next large spin state with $S = N/2-1$ has a reduced interaction energy gain $-I (N/2 - 1)(N/2)$, and so on. Thus typical energy spacings between collective states are given by $I \cdot N$, scaling with the number of TLSs. While in this paper we consider $I>0$, which is a result of second order perturbation theory for $\Delta>0$ and yielding large spin ground state, let us remark that in the opposite $I<0$ case, the ground state corresponds to the \emph{minimal} possible total spin $S=|S^z|$ for given $S^z$.

There are generically multiple energy-degenerate states with the same values of $S$ and $S^z$ that can be formed out of $N$ TLSs. This number is given by~\cite{Kaplan}
\be
\label{lambda}
f_S(N) = \frac{N! (2S+1)}{(\frac{N}{2}+S+1)!(\frac{N}{2}-S)!}.
\ee
We label these states by $t=1,..., f_S(N)$ and thus a general state is labeled as $| \mathcal{S} \rangle = |S,S^z,t \rangle$. There is a single large spin state $f_{N/2}(N)=1$, and $N-1$ states with $S = N/2-1$, and so on.

If $I=0$, the states with different $S$ are energy-degenerate, and can be transformed to another set of energy-degenerate states, corresponding to defined individual spin and vibrational states of each molecule. However, the energy-splitting due to finite $I$ (see Fig.~\ref{Lipkinlevels}) invalidates such transformation, and the individual states become undefined. Similar effect of spin-independent coordinate-dependent interactions between particles has been noticed already by Heitler \cite{Heitler27}.

We will see below a protection of the large spin states against influence of decoherence.

\subsection{Relation to the Dicke model}
\label{se:Dicke}
Model~(\ref{Lipkin}) can be derived from the Dicke model
\bea
\label{eq:Dicke}
H_D=\sum_k \nu_k a^\dagger_k a_k + \omega S^z + \sum_k( \gamma a_k S^+ + h.c.),
\eea
which is just Eq.~(\ref{eq:Hcavity}) in the limit where all the molecules under consideration are at the same point. As we now describe, the LMG model Eq.~(\ref{Lipkin}) is obtained for large enough $\epsilon_g$, where the photons can be "integrated out" and lead to an effective interaction between the $N$ TLSs.

\begin{figure}
\centering
\includegraphics*[width=.8\columnwidth]{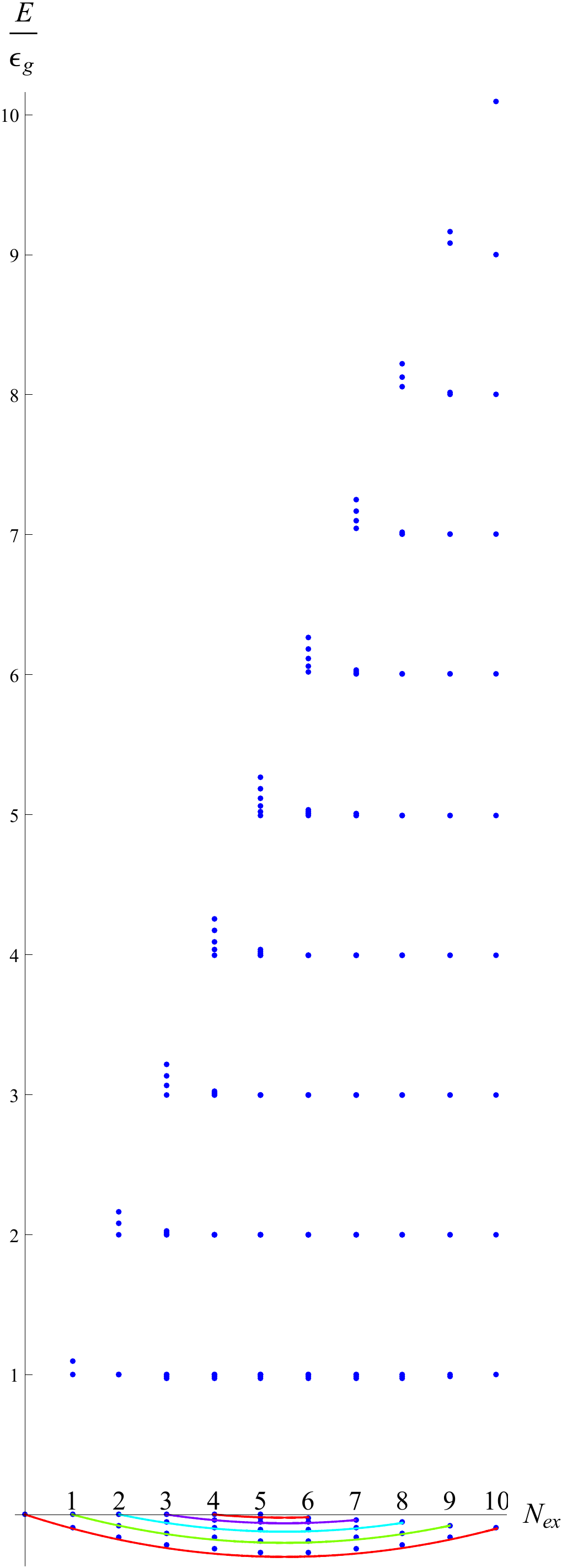}
\caption{(Color online) Full energy spectrum of the single mode Dicke model (\ref{eq:Dicke}) obtained by exact diagonalization, with $\epsilon_g=1$, $\omega=0$, $N=10$, and $\gamma  = 0.2$. One can see that the $n-$photon states are approximately quantized in units of $\epsilon_g$. The $n=0$ states are repelled to a negative energy. Fit of the zero-photon states to the LMG model (\ref{Lipkin}) with $I=\frac{\gamma^2}{ \epsilon_g}$ is shown by continuous parabolic lines (different colors correspond to different values of $S$ as in Fig.~\ref{Lipkinlevels}). }
\label{Dickelevels}
\end{figure}

Indeed, the Dicke Hamiltonian commutes with (i) the total number of excitations $N_{ex} = \sum_k a^\dagger_k a_k + \sum_i \frac{\sigma^z_i+1}{2}$ and with (ii) the total spin operator ${\bf{S}}^2 = S(S+1)$. For every value of $(N_{ex},S)$, as long as $N_{ex} \le N$, there can be $n_{ph}=0,1,...,N_{ex}$ photons, with the energy cost $\ge n_{ph} \epsilon_g$. When $\epsilon_g > \omega$, and $\epsilon_g - \omega \gg k_B T$,
only the zero photon $n_{ph}=0$ states survive in the low energy limit.  Then $N_{ex}=S^z+\frac{N}{2}$ ($n_{ph}=0$).
However, one can gain energy from virtual creation and annihilation of photons. The transition amplitude from a state with $S^z$ to $S^z -1$ via emission of a virtual photon involves the well known factor
\be
\label{Dickefactor}
S^-|S,S^z \rangle =\sqrt{S(S+1)-S^z (S^z-1)} |S,S^z-1 \rangle.
\ee
Hence, in second order perturbation theory we obtain a correction to the energy
\be
\delta E = -I [S(S+1)-S^z (S^z-1)]+ \mathcal{O} (\gamma^4),
\ee
which is just the LMG model with
\be
\label{eq:I}
I= \sum_k \frac{\gamma^2}{\nu_k - \omega} .
 \ee
The emergence of the LMG model as the low energy limit of the Dicke model is illustrated in Fig.~\ref{Dickelevels}.

We mention that the Dicke model displays a phase transition with spontaneously excited photons~\cite{Hepp}. Here we will not discuss this superradiant state. The absence of this phase transition is guaranteed (i) at zero temperature for $N \gamma_0^2 < \omega \epsilon_g$, where $\gamma_0^2=\sum_k \frac{\epsilon_g}{\nu_k} \gamma^2$, or otherwise (ii) by $T>T_c$ where $T_c$ is given by ${\rm{tanh}}  \frac{\omega}{2k_BT_c}=\frac{\epsilon_g \omega}{\gamma_0^2}$~\cite{Hepp}. However, even in the superradiant phase, there are virtual photons that will mediate the interaction between TLSs.

\subsection{Two-state vibration toy model}
\label{vibdiscrete}
Before moving to a study of the effects of decoherence in the next chapter, we here incorporate effects of \emph{discrete} vibrational modes within a simplified model amenable to exact diagonalization for small systems. In this model we keep only two vibrational states per molecule labeled by $\tau^z_i = \pm 1$, generalizing the Dicke model Eq.~(\ref{eq:Dicke}) to
\bea
\label{eq:DickeVib}
H_{D}'=\omega S^z + \sum_{i=0}^N (\epsilon \tau^z_i+C \sigma^z_i \tau^x_i) \nonumber \\
+(\sum_k \gamma  a_k S^+ +h.c.)+ \sum_k \nu_k a^\dagger_k a_k.
\eea
We now use this tractable model to test the fate of the collective levels in the presence of coupling to discrete vibrational modes.

After integrating out the photons, assuming large detuning $\Delta = \rm{min}_k \nu_k - \omega \gg \epsilon , C$, we obtain
\bea
\label{Hvibeff}
H_{{\rm{eff}}} &=&\omega S^z+ \sum_i (\epsilon \tau^z_i+C \sigma^z_i \tau^x_i)+H_{int}, \\
 H_{{\rm{int}}}&=&-I \sum_{ij} \sigma^+_j \sigma^-_i,\nonumber
\eea
where $I$ is given in Eq.~(\ref{eq:I}). Conventionally,  one first diagonalizes the local vibration Hamiltonian $H_{\tau} (\sigma^z) = \epsilon \tau^z +C \sigma^z \tau^x$ for each state of the electronic TLS $\sigma^z =\pm 1$. Thus, we may define two bases for the vibrational Hilbert space as eigenstates of $H_{\tau}(\sigma^z)$ for either value of $\sigma^z = \pm 1$:
\bea
\label{vibbasis}
{\rm{basis~1:}}~~~H_{\tau}(1) | \pm_{\uparrow} \rangle &=& \pm \sqrt{\epsilon^2+C^2} | \pm_{\uparrow} \rangle, \nonumber \\
{\rm{basis~2:}}~~~H_{\tau}(-1) | \pm_{\downarrow} \rangle &=& \pm \sqrt{\epsilon^2+C^2} | \pm_{\downarrow} \rangle.
\eea
Then, while the local part of the Hamiltonian is diagonal, the interaction creates vibrational transitions
\bea
\label{Iprime}
H_{int}=-I \sum_{i,j}   \sigma^+_j \sigma^-_i \times \nonumber \\
\left(
                                         \begin{array}{cc}
                                            |+_\uparrow \rangle &  |-_\uparrow \rangle \\
                                         \end{array}
                                       \right)_j
 \left(
  \begin{array}{cc}
    \langle +_\uparrow | +_\downarrow \rangle  & \langle +_\uparrow | -_\downarrow \rangle  \\
    \langle -_\uparrow | +_\downarrow \rangle  & \langle -_\uparrow | -_\downarrow \rangle  \\
  \end{array}
\right)
\left(
  \begin{array}{c}
    \langle +_\downarrow | \\
    \langle -_\downarrow | \\
  \end{array}
\right)_j \nonumber \\
  \times \left(
                                         \begin{array}{cc}
                                            |+_\downarrow \rangle &  |-_\downarrow \rangle \\
                                         \end{array}
                                       \right)_i
 \left(
  \begin{array}{cc}
    \langle +_\downarrow | +_\uparrow \rangle  & \langle +_\downarrow | -_\uparrow \rangle  \\
    \langle -_\downarrow | +_\uparrow \rangle  & \langle -_\downarrow | -_\uparrow \rangle  \\
  \end{array}
\right)
\left(
  \begin{array}{c}
    \langle +_\uparrow | \\
    \langle -_\uparrow | \\
  \end{array}
\right)_i.
  \eea
Explicitly the matrix elements are
\be
\label{matrixele}
 \left(
  \begin{array}{cc}
    \langle +_\uparrow | +_\downarrow \rangle  & \langle +_\uparrow | -_\downarrow \rangle  \\
    \langle -_\uparrow | +_\downarrow \rangle  & \langle -_\uparrow | -_\downarrow \rangle  \\
  \end{array}
\right)= \left(
  \begin{array}{cc}
   \cos(\alpha) & \sin(\alpha)  \\
    -\sin(\alpha)  & -\cos(\alpha)  \\
  \end{array}
\right),
\ee
where $\tan(\alpha) = C/\epsilon$. We see that in the local eigenbasis, the interaction $ \sigma^+_j \sigma^-_i$, which flips two TLSs at molecules $i,j$, also creates transitions in the vibrational states. The squares of the matrix elements (\ref{matrixele}) are our two-state model version of Franck-Condon factors. This leads to eigentates of the full system in which vibrations and electronic TLSs are generally entangled.

However, this coupling between TLSs and vibrations can be strongly suppressed in the regime of dominating interactions. Assuming that the interaction $I$ is large enough, we can neglect mixing of states with different $S$. If we are in the lower energy large $S=N/2$ spin state, which is permutation symmetric, we can replace the operator $\sigma^z_i$ in the interaction term $\propto C$ in Eq.~(\ref{Hvibeff}) by
 \be
\sigma^z_i \to 2S^z/N
 \ee
(see Sec. \ref{SecVibrSpectr} below).
Then for the large spin state the Hamiltonian becomes independent of the spin states of individual molecules
 \bea
\label{eq:decoupling}
H & \to &  H_S(S,S^z) + \sum_i (\epsilon \tau^z_i +\frac{2 S^z}{N}C \tau^x_i) =H_S(S,S^z)  \nonumber \\
&+&   \sqrt{\epsilon^2+\left( \frac{2 S^z}{N} C\right)^2}   \times \nonumber \\
\sum_i &[&  |  +^{S^z}(i)\rangle \langle  +^{S^z}(i) |- |  -^{S^z}(i)\rangle \langle  -^{S^z}(i) | ].
\eea
Here the vibrational eigenfunctions $|\pm^{S^z}\rangle$ and their eigenenergies depend only on the total many-body spin projection, unlike the eigenstates in Eq. (\ref{vibbasis}) dependent on the individual spins.
In the vibrational ground state all the molecules are in the $|  -^{S^z}\rangle$ state. The first vibrational excitation of the full system corresponds to exciting one molecule to the state $|  +^{S^z}\rangle$.

\begin{figure}
\centering
\includegraphics*[width=1\columnwidth]{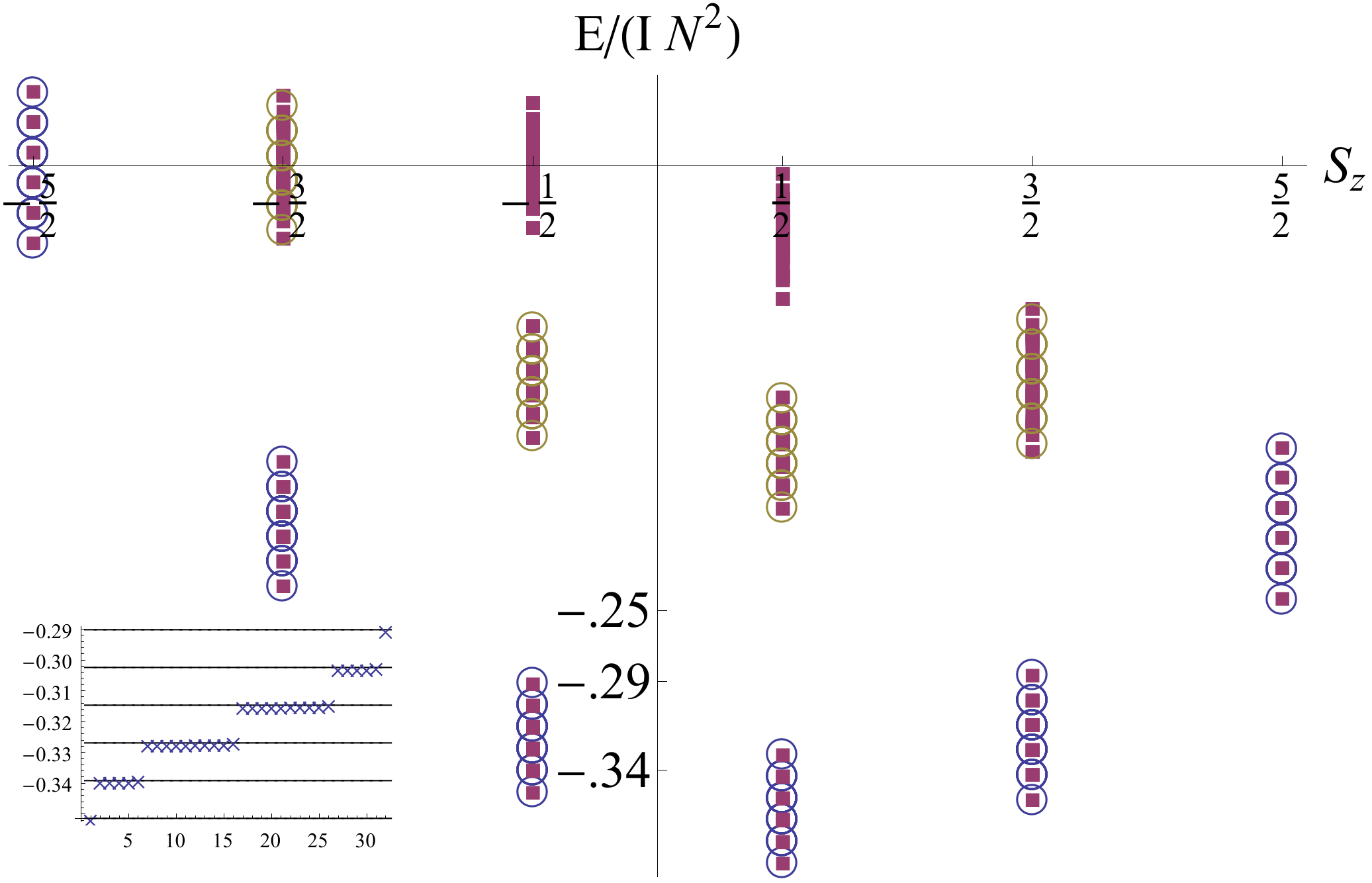}
\caption{(Color online) Spectrum of model (\ref{eq:DickeVib}) for $N=5$, $I=1$, $\epsilon=C=.15$ obtained by exact diagonalization (squares). Each level of the LMG model as in Fig.~\ref{Lipkinlevels} splits to $2^N$ vibrational states. Circles are fit to Eq.~(\ref{eq:decoupling}). The inset shows the $32$ energy levels corresponding to $S_z=-1/2$, $S=5/2$, versus Eq.~(\ref{eq:decoupling}).
}
\label{Dickelevelsvib1}
\end{figure}

This separation of the spectrum into a sum of decoupled TLSs part and vibration part that depends only on $S^z$ is confirmed by an exact diagonalization of the model (\ref{eq:DickeVib}) shown in Fig.~\ref{Dickelevelsvib1}. We can see that it matches the energy levels of the large spin $S = N/2$ manifold as well as smaller spin states. This decoupling is not exact, it relies on the formation of large spin, which is justified for large $IN \gg C$.

We will return to this simplified model in Sec.~\ref{se:PhysicalConsequence} to explicitly demonstrate the suppression of vibrational satellites in the emission spectrum.

\section{Decay rates of collective states}
\label{se:environment}
To study how the many-body states are influenced by coupling to a \emph{continuum }of bath modes, we integrate out the latter and obtain an effective theory of the spin system. We keep referring to these bosonic modes as ``vibrations" though their origin can be different as in general spin-boson models~\cite{Leggett87,Weiss99}. 

We use the resolvent formulation whose poles give the spectrum of the system.
Expanding the resolvent of the full system $R(z) = (z-H)^{-1}$ in powers of $V$, and tracing over the vibrations, one obtains (for details see for example Ref.~\onlinecite{Hewson})
\be
\label{Resolvent}
R(z) = \frac{1}{z - H_S - \Sigma(z)},
\ee
with the self-energy
\be
\label{Rsigma}
\Sigma(z) = \sum_{\{v\}} n_B(E_{\{v\}})   V (z+E_{\{v\}}-H_0)^{-1} V.
\ee
Here $n_B(E)\propto e^{-E/(k_BT)}$ is the normalized Boltzmann distribution. 
The imaginary part of $\Sigma(z)$ gives the Fermi's golden rule transition rate between different spin states mediated by exchange of a vibration quanta. We denote
\be
\label{GammaS}
 \Gamma_{\mathcal{S}}= - {\rm{Im}} \Sigma(z=E_{S, S^z}+ i \delta)
\ee
as the inverse half life time of the spin state $|\mathcal{S} \rangle$. As discussed in more detail in Appendix~A, the selection rules, mediated by $V$, allow exclusively for $S \to S, S \pm 1$ transitions. They  provide three terms in the decay rate of the collective state,
\bea
\label{result}
&&\Gamma_{S,S^z}= \frac{2 \pi}{2S+1}   \sum_{S'}  \times \\
&&[ \delta_{S',S} \left( (N/2+S+1)   \frac{{S^z}^2}{S}+ ( N/2-S)  \frac{{S^z}^2}{S+1} \right)  \mathcal{A}(0) \nonumber \\
&+& \delta_{S',S+1} (N/2-S)   \frac{(S+1)^2-{S^z}^2}{S+1}  \mathcal{A}(E_{S+1, S^z}-E_{S, S^z}) \nonumber \\
&+&  \delta_{S',S-1} (N/2+S+1)  \frac{S^2-{S^z}^2}{S}  \mathcal{A}(E_{S-1, S^z}-E_{S ,S^z})]. \nonumber
\eea
In the case of $N=1$, $S=S'=S^z=1/2$, this equation gives the level width of a single molecule~\cite{Skinner86}
\be
\Gamma_{1/2,1/2}= \pi \mathcal{A}(0)=\frac{1}{2T_2}.
\ee
Then the  spectral function $\mathcal{A}(0)$ is related to the single-molecule relaxation time $T_2$.

Equation (\ref{result}) is the main result of this section. We now discuss its content. The terms $\propto \delta_{S', S \pm 1}$ correspond to transitions $S \to S' \ne S$. Starting from a large spin state $S \sim N/2$, and for a nearly unpolarized state $|S^z| \ll S$, we see from the overall coefficients that transitions $S \to S+1$ are suppressed by a factor $(\frac{N}{2}-S)$ (which exactly vanishes for $S=N/2$), while transitions $S \to S-1$ have an increased rate $\propto N$, which is the phase space corresponding to the number of possible final states with smaller spin, Eq.~(\ref{lambda}). However, transitions from the large spin state require a finite energy to be extracted from the vibrational baths. The thermal energy $k_B T$ will be exceeded by the required energy difference $E_{S-1, S^z}- E_{S ,S^z} = I N$ for large enough $N$. Thus $\mathcal{A}(E) \propto n_B(E) \to 0$ for
\be
\label{temperature}
I N \gg k_B T.
\ee
Then the large spin state becomes \emph{stable}.

The most prominent regime to study this large spin state, is near the unpolarized state where $(S^z)^2 \le N$. As can be seen in Fig.~2, this corresponds to the lowest energy window of size $\delta E  = (I \cdot N)$, which includes only the large $S=N/2$ manifold, with  $\sim \sqrt{N}$ states.

In this large interaction regime only the terms $\propto \mathcal{A}(0)$ in Eq.~(\ref{result}), leaving $S$ fixed, contribute. These terms, however, are of order $\frac{(S^z)^2}{S}$, namely are suppressed for large $N$. Thus
\be
\label{GammaoneoverN}
\Gamma_{S,S^z} \cong \frac{1}{N T_2} ,~~~(S = N/2,S^z =\mathcal{O}(1)).
\ee
Thus, the large spin state enjoys from a $1/N$ reduction of the dephasing rate.

We note that the result Eq.~(\ref{result}) is not valid for $I=0$ since it assumes initial and final \emph{collective} states,  while for $I=0$ processes of decoherence happen within a single molecule or in its near vicinity. Self-consistenly, to ensure the stabilization of collective states we demand $I N \gg \Gamma_{S,S^z}$.

The real part of the self-energy $\Sigma = \Sigma' + i\Sigma''$ provides information on energy shifts of the spin states due to their coupling to the vibrations. In Appendix~B we estimate these corrections and find that they are subdominant, namely they are of order $\mathcal{O}(1)$, as compared to the $\mathcal{O}(N)$ energy difference between collective states.


\section{Molecular Emission Spectrum}
\label{se:PhysicalConsequence}
We now discuss physical signatures of the collective states in the emission spectrum. For comparison, for a single TLS the emission spectrum has a Lorentzian line shape with $W(E) \propto [E- (\omega+\Delta \omega)+(1/T_{2})^2]^{-1}$
where $\Delta \omega$ is an energy shift
and $T_2$ results from decoherence. Both the position and width of the peak are modified in a system of $N$ TLSs, see for example related studies involving plasmons or polaritons~\cite{Ziolkowski95,Sukharev11}. Here, we will identify the role of interactions and pinpoint how these effects in the emission spectrum scale with $N$.

For simplicity consider an identical coupling of the $N$ TLSs to classical light,
\be
\delta H = \Omega \sum_{i} \sigma^-_i e^{i E t}+h.c. = \Omega S^- e^{i E t}+h.c..
\label{deltaH}
 \ee
Notice that we are implicitly distinguishing the emitted photons from the cavity photons mediating the interaction. The latter are emitted and absorbed multiple times, which is assisted by the cavity. On the contrary the emitted photons contributing to the emission spectrum $W(E)$ propagate in free space, and yet have a finite coupling to the TLSs inside the cavity.

 Emission occurs via transitions $S^z \to S^z-1$. Consider the system in an initial state $| i \rangle = |S,S^z,t \rangle$. As shown in Appendix C, the transition rate to final state $| f \rangle=|S,S^z-1,t' \rangle$ is proportional to
\bea
\label{gammaif}
W(E) \propto \Omega^2 |\langle f |S^- | i \rangle|^2 \times \nonumber \\
\frac{\Gamma_i + \Gamma_f}{(\omega + 2I(S^z-1) + \Sigma_{i}'- \Sigma_f'-E)^2 + (\Gamma_i+\Gamma_f)^2}.
\eea
The level widths $\Gamma_{i,j}$ and the level shifts $\Sigma_{i,f}'$  were introduced in the previous section; it is assumed that the width is dominated by the vibrational modes rather than by the coupling $\Omega$ to the emitted light, namely $\Gamma_{i,f} \gg \Omega$. Few effects apparent in Eq.~(\ref{gammaif}) should be noted:

(i) The Dicke factor  (see Eq.~\ref{Dickefactor}) that strongly depends on $S^z$:
\be
|\langle f |S^- | i \rangle|^2 \propto \begin{cases}
  N,  & S^z \sim N/2,\\
  N^2, & S^z=\mathcal{O}(1).
\end{cases}
\ee
This Dicke enhancement factor is \emph{independent} of the interactions between the TLSs.

(ii) The energy of the transition $(S,S^z) \to (S,S^z-1)$,
\be
\label{emissionenergy}
\omega_{S^z}=E_{S ,S^z} -E_{S, S^z-1}= \omega+ 2I(S^z-1),
\ee
is strongly shifted by the interactions depending on the polarization state $S^z$. Focusing on the lowest energy window $\delta E = NI$ in Fig.~\ref{Lipkinlevels}, with $|S^z| < \mathcal{O}(\sqrt{N})$, this shift is of order $\sqrt{N}$. It overcomes the level shifts $\Sigma_{i,f}'$ of order unity. 

\begin{figure}
\centering
\includegraphics*[width=1\columnwidth]{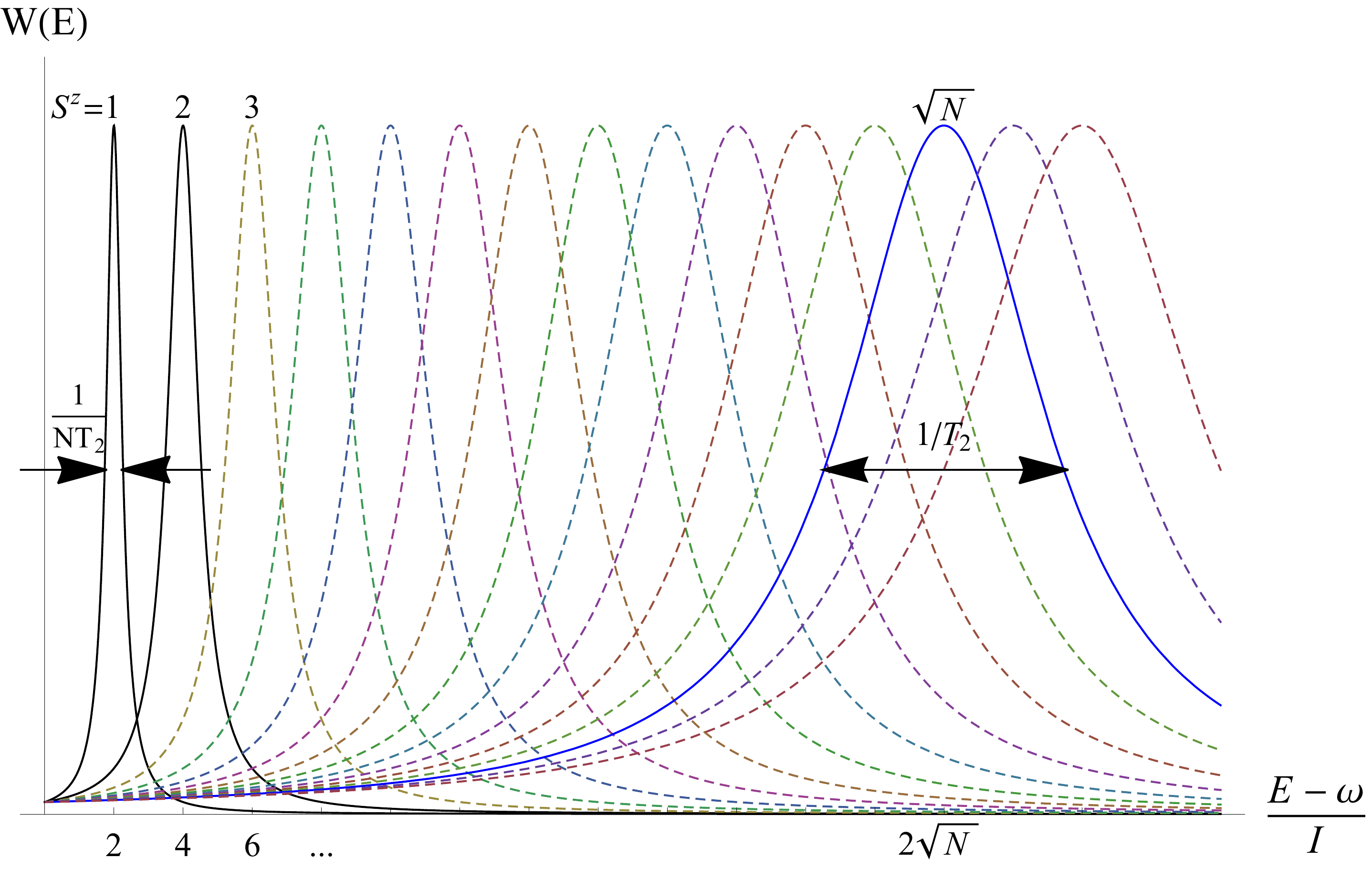}
\caption{(Color online) Schematic depiction of the emission spectrum of $N$ interacting TLSs. Different peaks correspond to different initial values of $S^z$ with the peak positions given in Eq.~(\ref{emissionenergy}) up to $\mathcal{O}(1)$ additional level shifts. $\omega_p$ is a polaronic shift of the ground state energy in the $S^z$ and $S^z-1$ states.
In the large $S$ regime
the level width increases with $S^z$ as $\Gamma \sim [(S^z)^2/S] \times 1/ T_2$, reaching values of order $1/T_2$ for $S^z \sim N^{1/2}$, and values of order $1/(N T_2)$ for $S^z \ll S$. }
\label{fg:Emission}
\end{figure}

(iii) As follows from Eq.~(\ref{GammaoneoverN}) the width of the emission line $\Gamma_i + \Gamma_f$ is strongly suppressed. Specifically, the peak width is $N$-times sharper than the single molecule line width $1/T_2$ for the $S^z = \mathcal{O}(1)$ and becomes of order $1/T_2$  for $S^z \sim \sqrt{N}$. This is depicted schematically in Fig.~\ref{fg:Emission}. 

    Along the cascade of decay due to emission, $S^z \to S^z-1 \to S^z-2$ which can be analyzed \emph{e.g.} solving rate equations, the system is in a probabilistic superposition of collective states with different $S^z$. Then the emission spectrum also contains a superposition of peaks whose positions and widths are shown in Fig.~\ref{fg:Emission}. They will become well isolated if $I N \gg \frac{1}{T_2}$. The relative peak heights in Fig.~\ref{fg:Emission} along this cascade, was not calculated here.

\subsection{Vibrational structure of the emission spectra}\label{SecVibrSpectr}
Transitions between discrete vibrational levels typically lead to additional peaks in the emission spectrum, here referred to as Franck-Condon ``satellites". We have ignored these satellites in the previous subsection. However we now demonstrate that this was done with a good reason: in the large-interaction limit these satellites are suppressed as $1/N$.

The emission spectrum
\be
\label{GFR}
W(E)= \sum_{E_{f i}}a(E_{f i})\delta(E +E_{f i})
\ee
contains several peaks. The peak intensities $a(E_{f i})$
are determined by the matrix elements of the interaction (\ref{deltaH}) with the classical field as
\be
a(E_f - E_i)= 2 \pi \Omega^2 |\langle \Psi_f | S^-| \Psi_i \rangle|^2 .
\ee
Here $| \Psi_i \rangle$ and $| \Psi_f \rangle$ are eigenstates of the Hamiltonian (\ref{eq:Hmodel}), corresponding to the total spin projections $S^z$ and $S^z-1$, respectively, and  $E_i$ and $E_f$ are their eigenenergies.

\begin{figure}
\centering
\includegraphics*[width=.8\columnwidth]{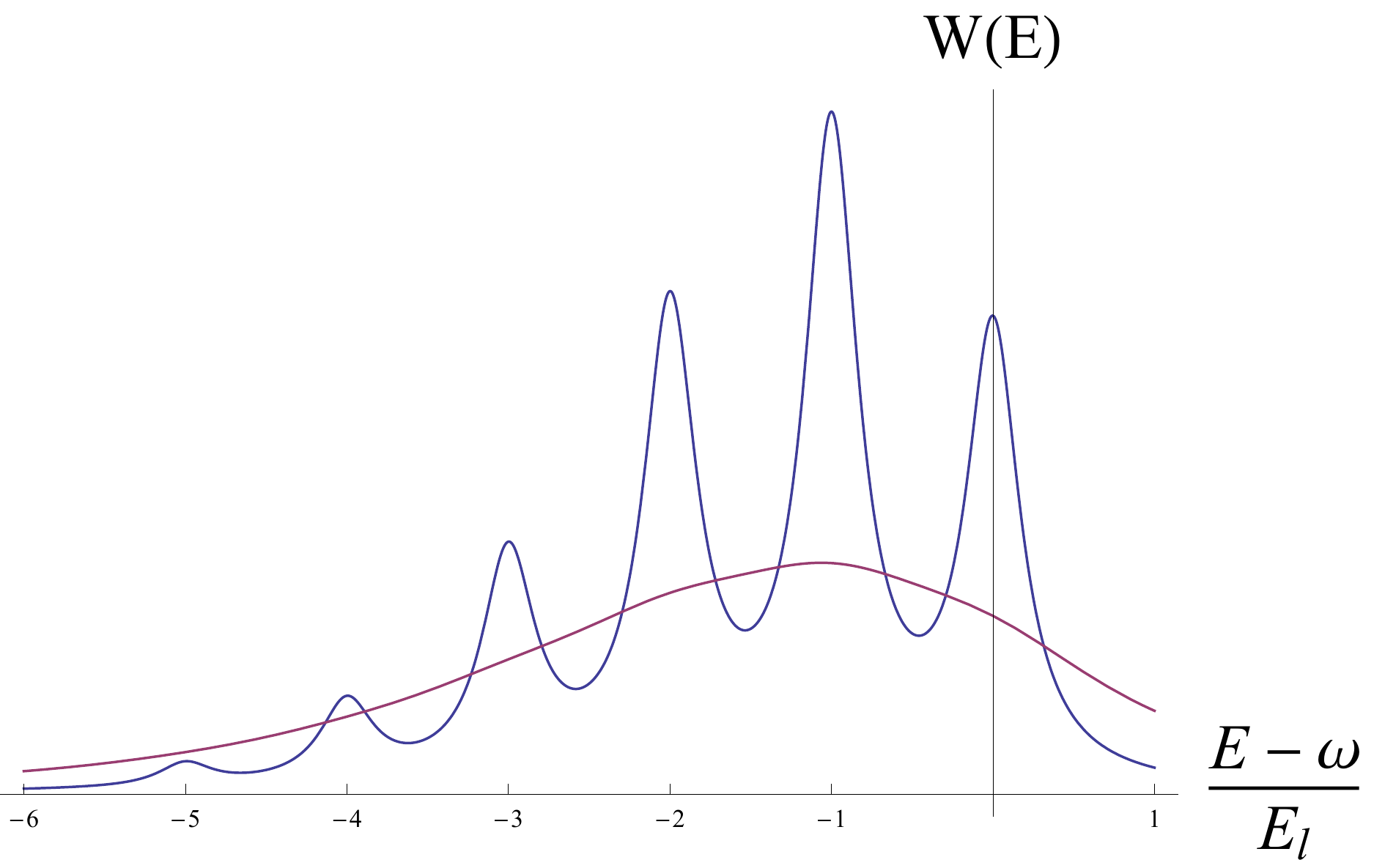}
\caption{(Color online) Schematic depiction of the emission spectrum of one molecule including (i) vibrational satellites (for one mode $l$) according to the Franck-Condon factor Eq.~(\ref{eq:FCf}), and (ii) additional broadening $1/T_2 = 0.5 E_l$ and $0.1 E_l$. Here $C_l/E_l=0.6$.}
\label{fg:FC1}
\end{figure}

For one molecule, vibrational transitions occur due to the linear coupling $C_l$ in Eq.~(\ref{eq:Hmodel}) which we write for clarity for one mode of energy $E_l$ as
\bea
H_{vib,\uparrow} = E_l v_{l,i}^\dagger v_{l,i} + (C_l  v_{l,i}+h.c.), \nonumber \\
H_{vib,\downarrow} = H_{vib,\uparrow} -2C_l (  v_{l,i}+h.c.).
\eea
The transition from the vibrational ground state $| 0_\uparrow \rangle$ to a final excited vibrational state $| n_\downarrow \rangle$ ($n\ge 0$) corresponds to the peak with the intensity $a (\omega - n E_l)= 2 \pi \Omega^2 F_{n}(C)$, where the Franck-Condon factor is explicitly given by
\be
\label{eq:FCf}
F_{n}(C)=|\langle n_\downarrow | 0_\uparrow \rangle|^2 = \frac{1}{n!}\left( \frac{2C_l}{E_l}\right)^{2n}\exp\left(-\frac{4 C_l^2}{E_l^2}\right).
\ee
The resulting emission lines are shown in Fig.~\ref{fg:FC1}.

\begin{figure}
\centering
\includegraphics*[width=.8\columnwidth]{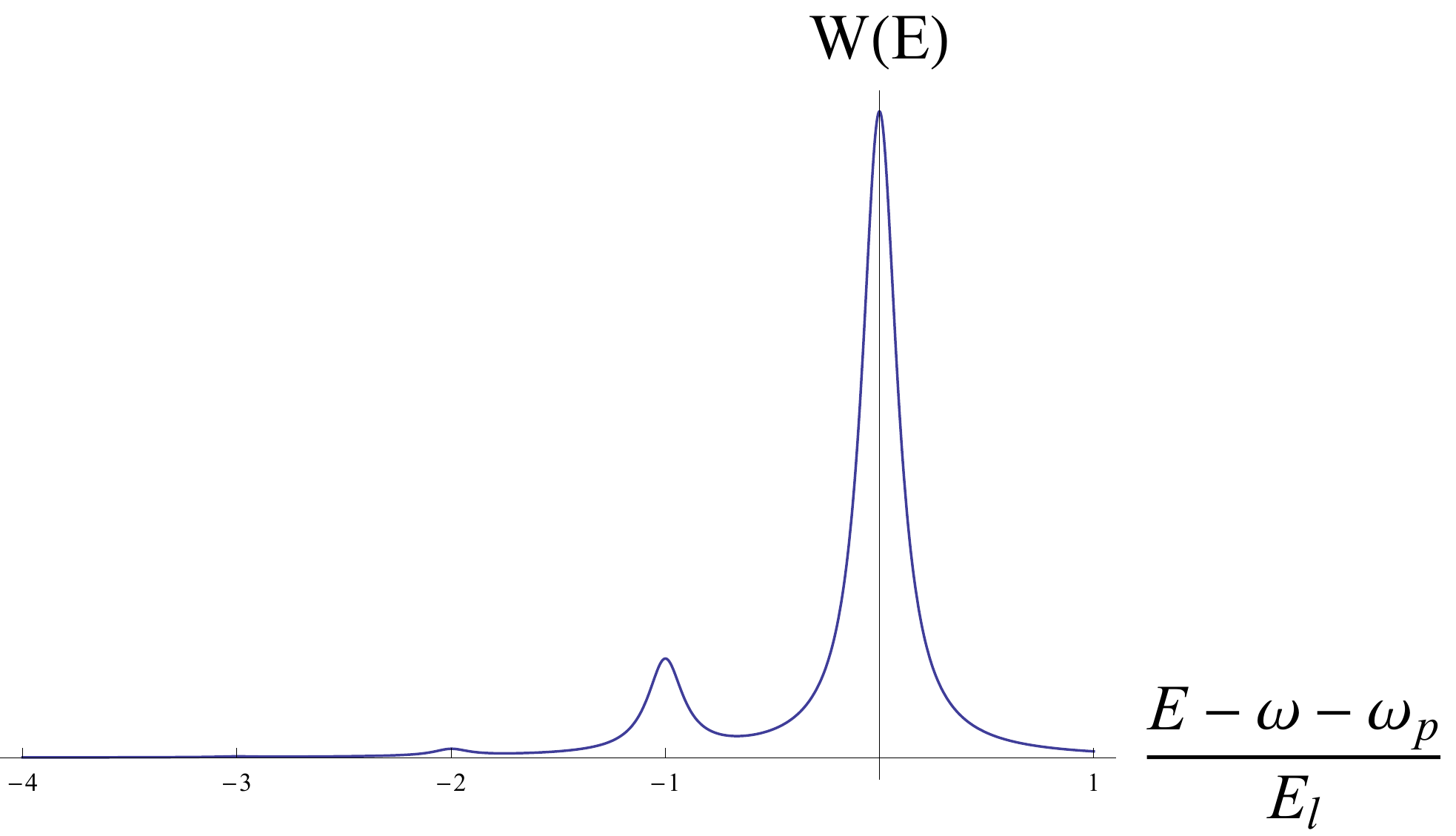}
\caption{(Color online) Schematic depiction of the emission spectrum as in Fig.~\ref{fg:FC1} but for $N=10$ molecules. Here $1/T_2 = 0.5E_l$ and $\omega_p$ is an additional polaronic shift.}
\label{fg:FC2}
\end{figure}

For many molecules the Hamiltonian (\ref{eq:Hmodel}) commutes with $S^z$ and all its terms, except for $V$, commute with $S$. In the limit of large $I \cdot N$, the large energy gaps between states with different $S$ (see Fig. \ref{Lipkinlevels}) allow us to neglect their coupling and consider the projection of the Hamiltonian (\ref{eq:Hmodel}) to the states $|S,S^z,t \rangle$
\bea
H_{vib,S,S^z} = \sum_t |S,S^z,t \rangle \langle S,S^z,t|~(H_{v} +V) \nonumber \\
\sum_{t'} |S,S^z,t' \rangle \langle S,S^z,t'|
\eea
(the contribution of $H_S$ is omitted here as it can lead to an energy shift only). Suppose now that $N$ TLSs form a large spin $S = N/2$ state. As this state $|N/2,S^z\rangle$ is symmetric with respect to permutations of the molecular spins,
\be
\langle N/2,S^z|\sigma^z_i|N/2,S^z\rangle = \frac{1}{N}\langle N/2,S^z|\sum_i \sigma^z_i|N/2,S^z\rangle .
\ee
Then the vibration Hamiltonian becomes
\be
H_{vib,N/2,S^z} \to \sum_i  E_l v_{l,i}^\dagger v_{l,i} + \frac{2S^z}{N} (C_l  v_{l,i}+h.c.).
\ee
One can immediately see that the shift of the vibrational potential in the process $S^z \to S^z-1$ is reduced by a factor $1/N$,
\be
H_{vib,N/2,S^z-1} = H_{vib,N/2,S^z}-\frac{2 C_l}{N} (  v_{l,i}+h.c.).
\ee
This leads to dramatic effects in the emission spectrum. The main peak corresponds to the process $|0_\uparrow,0_\uparrow,0_\uparrow...\rangle \to |0_\downarrow,0_\downarrow,0_\downarrow...\rangle$ and has an intensity
\be
a(\omega_{S^z})=2 \pi \Omega^2 (\frac{N}{2}+S^z)(\frac{N}{2}-S^z+1) [F_{0}(C/N)]^N.
\ee
The first satellite corresponds to a final state with an $n=1$ excitation in one of the $N$ molecules with intensity
\bea
a(\omega_{S^z}-E_l)=2 \pi \Omega^2 (\frac{N}{2}+S^z)(\frac{N}{2}-S^z+1) \times \nonumber \\ N  [F_{0}(C/N)]^{N-1}  F_{1}(C/N).
\eea
Similarly the second and higher satellites can be obtained. The ratio of the first satellite intensity to that of the main peak is given by
\be
\frac{N F_{1}(C/N)}{F_{0}(C/N)} =\left( \frac{2C_l}{E_l}  \right)^2 \frac{1}{N}.
\ee
Hence we observe a $1/N$ reduction of the satellites. The emission lines based on this analysis are shown in Fig.~\ref{fg:FC2}.

In order to characterize the cumulative effect of the satellite suppression, let us introduce the total line intensity
\be
a=\int W(E) d E = \sum_f a(E_f - E_i)
\ee
The summation over $f$ can be expanded over the complete set of states $| \Psi_f \rangle$, since the matrix elements vanish for non-coupled states. Then
\be
a= 2 \pi \Omega^2 |\langle \Psi_i |S^+ S^-| \Psi_i \rangle|^2 =2 \pi \Omega^2 (\frac{N}{2}+S^z)(\frac{N}{2}-S^z+1)
\ee
and the ratio of the main peak intensity to the total one, given by
\be
\frac{a(\omega_{S^z})}{a}=[F_{0}(C/N)]^N=\exp\left(-\frac{4 C_l^2}{N E_l^2}\right),
\ee
tends to unity at large $N$.
Below we will confirm this effect in the two-state vibration model.

\subsubsection{Calculation for the two-state vibration model}

To demonstrate explicitly the picture described above for the satellites suppression, we consider the simplified model of discrete vibrational modes in Sec.~\ref{vibdiscrete}. The emission lines correspond to transitions between many body levels (see Fig.~\ref{Dickelevelsvib1}) where $S^z$ changes by 1. Consider starting from the initial state $|\Psi_i \rangle$ being the vibrational and electronic ground state with given $S^z$.
For one molecule the electronic transition from $S^z = 1/2$ to $S^z=-1/2$ (or $|\uparrow \rangle \to |\downarrow \rangle$) is associated with two emission peaks with intensities $a( \omega)$ and  $a( \omega - 2 \sqrt{C^2+\epsilon^2})$. The first peak corresponds to a transition from the vibrational ground state $| -_\uparrow \rangle$ [see Eq.~(\ref{vibbasis})] to the new ground state $| -_\downarrow \rangle$, with matrix element $a_0=\cos^2(\alpha)$ [see Eq.~(\ref{matrixele})]. The satellite peak corresponds to a transition to the excited vibrational state $| +_\downarrow \rangle$, with matrix element $a_1=\sin^2(\alpha)$.
Thus the ratio of the satellite to the main peak intensity is given by $\tan^2(\alpha)=C^2/\epsilon^2$.
We now explore how this ratio evolves for a few interacting molecules.

\begin{figure}
\centering
\includegraphics*[width=.95\columnwidth]{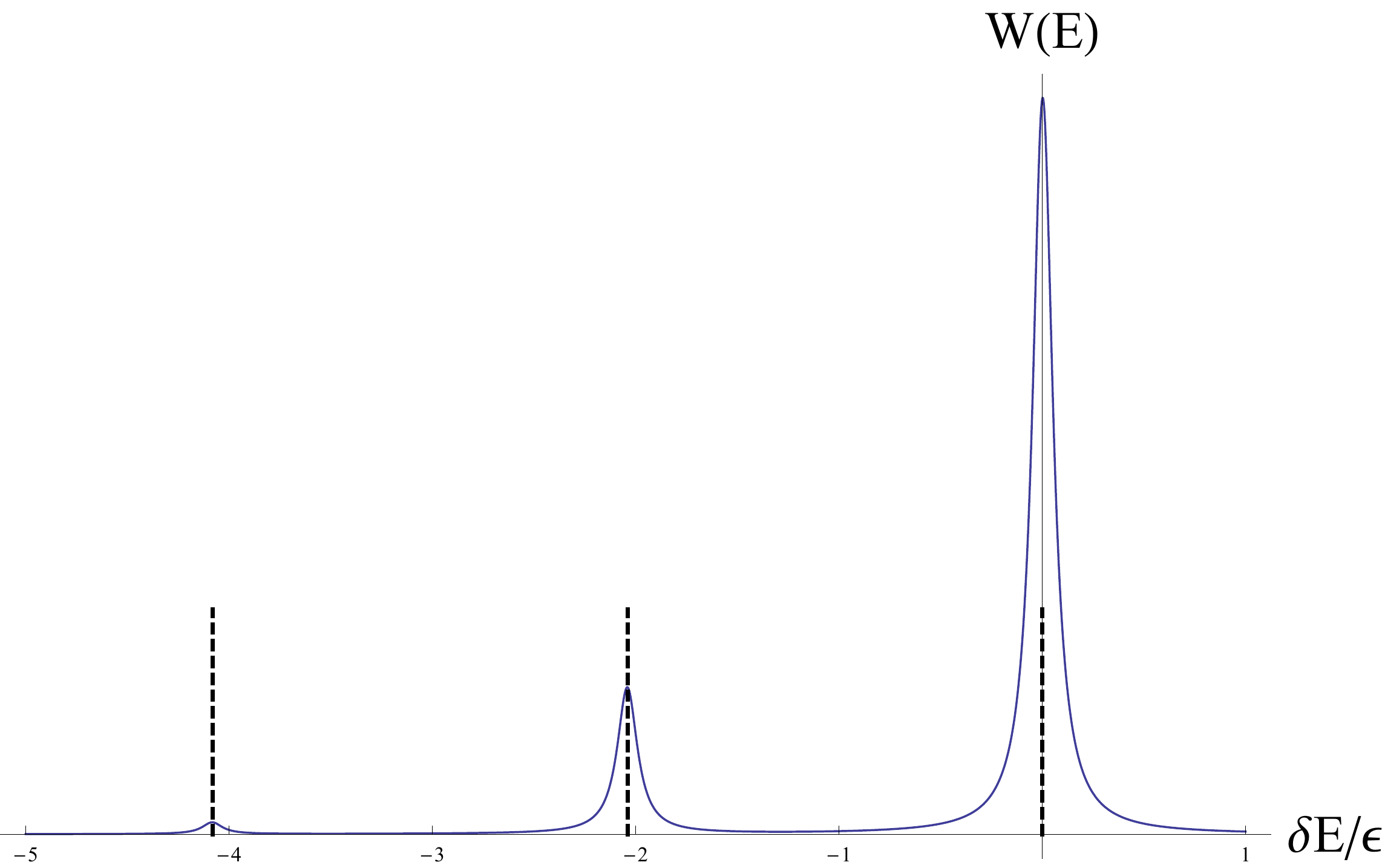}
\caption{(Color online) Emission spectrum for the model (\ref{eq:DickeVib}) with $N=5$, $\epsilon=C=0.01, I=1$. We start in the $S^z=1/2$ ground state (belonging to the total spin $5/2$ ``parabola" in Fig.~\ref{Dickelevelsvib1}). }
\label{fg:EmissionFC}
\end{figure}

As described in Sec.~\ref{vibdiscrete}, when $I$ is large enough, so that we are in the lowest energy large $S=N/2$ spin state which is symmetric, we can replace the operator $\sigma^z_i$ in the interaction term $\propto C$ in Eq.~(\ref{Hvibeff}) by $\sigma^z_i \to 2S^z/N$. The initial and final Hamiltonians of the vibrations are given by  $H_{\tau} (\sigma^z)$ with either $\sigma^z= 2S^z/N$ or $\sigma^z= 2(S^z-1)/N$, respectively. We start in the ground state of the configuration with the initial $S^z$ whose energy is $-N \sqrt{\epsilon^2+ \left(2 \frac{S^z}{N} C \right)^2}$ (all $N$ molecules in the  vibrational ground state $| -^{S^z}\rangle$). The main peak is obtained by going to the ground state of the new Hamiltonian with $S^z-1$ in all the molecules (all $N$ molecules in the new vibrational ground state $| -^{S^z-1}\rangle$), having energy $-N \sqrt{\epsilon^2+ \left(2 \frac{S^z-1}{N} C \right)^2}$. Thus the main emission peak occurs at the energy
 \be
 E_{main} = H_{int}(S,S^z) - H_{int}(S,S^z-1) + \delta E,
 \ee
where we have separated the vibrational contribution
 \be
 \delta E = -N \sqrt{\epsilon^2+ \left(2 \frac{S^z}{N} C \right)^2} + N \sqrt{\epsilon^2+ \left(2 \frac{(S^z-1)}{N} C \right)^2}.\nonumber
 \ee
The $n$-th satellite corresponds to flipping $n$ molecules to their vibrational excited state $| +^{S^z-1}\rangle$, and corresponds to the emission line
  \be
 \delta E_n = -N \sqrt{\epsilon^2+ \left(2 \frac{S^z}{N} C \right)^2} + (N-2 n) \sqrt{\epsilon^2+ \left(2 \frac{S^z-1}{N} C \right)^2}.\nonumber
 \ee
This set of satellites and the main peak ($n=0$) is plotted as dashed lines in Fig.~\ref{fg:EmissionFC} and perfectly agrees with the position of the peaks obtained by computing the Fermi's golden rule Eq.~(\ref{GFR}) using the numerically obtained eigenstates for a small $(N=5)$ system with large interaction $I$.

In addition, the heights of the satellites are strongly suppressed. The height-ratio of the strongest satellite to the main peak is
\be
N \tan^2 \frac{\alpha_{S^z}-\alpha_{S^z-1}}{2}, ~~~\alpha_{S^z}=\arctan\frac{2 S^z C}{N \epsilon}.
\label{rata2st}
\ee
The factor of $N$ counts possible choices of the molecule, in which the excitation $| +^{S^z-1} \rangle$ is located in the final state, and the remaining function is the ratio of matrix elements $\frac{| \langle +^{S^z-1} |-^{S^z} \rangle |^2}{| \langle  -^{S^z-1}| -^{S^z} \rangle |^2}=\tan^2 \frac{\alpha_{S^z}-\alpha_{S^z-1}}{2}$. The ratio (\ref{rata2st}) perfectly agrees with the numerical results. Similarly, higher satellite peaks are suppressed by higher power of this same small factor.

For $S^z = N/2$ and large $N$ this ratio tends to $\frac{1}{N} \frac{(C/\epsilon)^2}{[1+(C/\epsilon)^2]^2}$. For $S^z /N \ll 1$ this ratio becomes $ \frac{1}{N} \frac{C^2}{\epsilon^2}$. In either regime we obtain the $1/N$ suppression of the satellites.

To conclude, we have shown a $1/N$ suppression of the vibrational satellites, which is a crucial effect of the collective interacting set of TLSs.

\section{Observability}
\label{se:obs}
We now discuss the observability of this collective effect in a microcavity filled with dye molecules, referring to the experimental parameters of Klaers \emph{et. al.}~\cite{Klaers10}.
The emission line of a single molecule has a typical frequency in the visible range $\omega \sim 500{\rm{THz}}$ and width of the order of $50{\rm{THz}}$. This width corresponds to many vibrational satellites broadened by $1/T_2$. The system is at room temperature $k_B T \sim 10 {\rm{THz}}$.

Following Ref.~\onlinecite{Sela14}, the interaction $I$ in our model can be estimated using the typical emission time from a single TLS in the cavity $\tau \sim 3 {\rm{nsec}}$. This rate is related via Fermi's golden rule to the product of the coupling constant $\gamma^2$ and the density of states $\frac{m}{2\pi}$,
\be
\frac{1}{\tau} =L^2 \gamma^2 m \sim  1 {\rm{GHz}}.
\ee
Using Eq.~(\ref{eq:IK}) we see that the interaction is determined by the same parameters,
\be
I \sim \frac{1}{\tau} \sim  1 {\rm{GHz}}.
\ee
This can be negligible compared to the decoherence, which is restricted by the width and can have a value up to  $1/T_2\sim50{\rm{THz}}$. However, could coherence be restored due to large $N$?
Using Eqs.~(\ref{eq:N}) and (\ref{eq:ell}), we obtain
\be
N \sim \left( \frac{c}{\epsilon_g} \right)^3 \frac{1}{\delta^3} \frac{1}{(\Delta / \epsilon_g)}\sim \left( \frac{d}{\delta} \right)^3 \frac{\epsilon_g}{\Delta}.
\ee
Since the separation between the mirrors $d=1.5 \mu m$ exceeds  the separation between molecules $\delta \sim 10nm$ by two orders of magnitude, and since the detuning is naturally much smaller than the cavity frequency $\epsilon_g = c \frac{n_z \pi}{d}~10^{15}{\rm{Hz}}$ (the standing wave number is $n_z=7$), we  expect large values of $N$, which will satisfy the condition $I \cdot N \gg 1/T_2, k_B T$.

While the experiment~\cite{Klaers10} concentrated on Bose-Einstein condensation of photons at room temperature, the present scenario requires
(i) large detuning $\Delta>0$ such as to push the cavity mode to high frequencies compared to the TLS transition, (ii) lower temperature leading to reduced $1/T_2$, and (iii) high polarization state (nearly half of the TLSs in the excited state).

Yet, our crude estimate of $N$ should be taken with a grain of salt, as it was obtained using crude assumptions of constant interaction strength $I_{ij}$, ignoring near field effects as well as fluctuations in its sign due the $\sin(z_i n_z \pi /d)$ factors in Eq.~(\ref{eq:gammaki}) and the $n_z>1$ condition in the experiment.

\section{Conclusions}
\label{se:conclusion}
We studied a competition between local decoherence and all-with-all interaction between $N$ TLSs.
The physical situation considered here, where each TLS contains a separate bath of oscillators, may be realized in an ensemble of dye molecules in an optical micro-cavity. The new many body physics studied here corresponds to a rather unexplored regime, with nearly equal population of excited and deexcited TLSs.

The effect of interaction between TLSs is a formation of a many-body states, which results \emph{e.g.} in strong shift of the emission line, level narrowing by $1/N$ as well as a $1/N$ suppression of vibrational satellites. 

Interactions, mediated by two-dimensional cavity modes, allow us also to control many-body states of two-level atoms, where the obstructing effects of molecular vibrations are absent.

We considered an ideal system with equal interaction between all TLSs, ignoring effects of disorder. A more generic model for the interaction Hamiltonian is $-  \sum_{i, j} I_{ij}  \sigma_i^+ \sigma_j^- $ with $I_{ij} \propto \gamma_i \gamma_j$ with  random $\{ \gamma_i\}$, \emph{e.g.}, due to the random $\{ z_i \}$ locations.  We speculate that the emergence of large spins states separated by a large energy $\propto (IN)$ from all other states is a robust effect that survives in the presence of disorder. Proving this assertion is left for a future study.

\section{Acknowledgments}
We thank Iacopo Carusotto, Emanuele Dalla Torre, Moshe Goldstein, Abraham Nitzan, and
Angelo Russomanno for discussions, and
acknowledge funding by the Israel Science Foundation
grant No.~1243/13 (E.S.), grant No.~1309/11 (V.F.) and by the Marie Curie CIG grant No.~618188 (E.S.). V.F. is grateful to PCS IBS, Daejeon, Korea for hospitality.

\appendix
\section{Calculation of the level width}
The Hamiltonian $H_0$ contains $H_S$ [see Eq. (\ref{eq:Hmodel})], which is independent of the molecular vibrational and translational coordinates, and $H_v$, which is independent of the electronic degrees of freedom, treated here as spins. Each of the parts is permutation-invariant. Therefore, the total many-body wavefunction is represented as (see \cite{Kaplan,yurovsky15}
\be
\label{tilPsi}
\tilde{\Psi}_{r\{v\}S_{z}}^{(S)}=f_{S}^{-1/2}\sum_{t}\tilde{\Phi}_{tr\{v\}}^{(S)}|S,S^z,t \rangle.
\ee
Here the spin $|S,S^z,t \rangle$ and spatial $\tilde{\Phi}_{tr\{v\}}^{(S)}$ wavefunctions form bases of irreducible representations of the symmetric group associated with the Young diagram $\lambda=[N/2+S,N/2-S]$. These representations have dimensions $f_S(N)$ [see Eq. (\ref{lambda})], and the labels of the basic functions $t$ are standard Young tableaux of the shape $\lambda$. For bosons, the representation of the spatial wavefunctions is associated with the same Young diagram $\lambda$, while for fermions it is associated with the conjugate diagram $[2^{N-S/2}, 1^S]$. This provides the correct bosonic of fermionic permutation symmetry of the total wavefunction. The Young tableaux $r$ label different representations, associated with the same Young diagram.

In the wavefunction (\ref{tilPsi}) the spatial wavefunctions are represented as symmetrized wavefunctions of non-interacting dye molecules in the vibrational-translational states $|v_i\rangle$ with energies $E_{v_i}$. It neglects two-body coordinate-dependent interactions between the molecules. In a thermal system, multiple occupation of a translational state has negligible probability. Thus all $v_i$ are different, although several molecules can be in the equal vibrational states.

States with different total spins can be mixed by interactions which depend on both spins and coordinates. In the present case, the relevant interaction has the form
\be
V=\sum_i \frac{1}{2} \sigma^z_i U(i) .
\ee
Here spin-independent $U(i)$ can describe both interactions with internal degrees of freedom, e.g., the linear coupling of Eq. (\ref{eq:Hmodel}), and arbitrary interactions with the environment dependent on the internal and translational coordinates of the molecule.

The lifetime of the many-body states is determined by the imaginary part (\ref{GammaS}) of the self-energy, calculated with the Fermi's golden rule
\begin{widetext}
\be
\label{ImSigma}
{\rm{Im}} \Sigma(z=E_{S S^z}+ i \delta)=-\pi\sum_{\{v\},\{v'\},r',S'}
\left |\langle \tilde{\Psi}_{r'\{v'\}S_{z}}^{(S')}|V|\tilde{\Psi}_{r\{v\}S_{z}}^{(S)}\rangle\right |^2
\delta(E_{S' ,S^z}-E_{S ,S^z}+\sum_i(E_{v'_i}-E_{v_i})).
\ee
The $S^z$-dependence of the matrix elements can be extracted by using the Wigner-Eckart theorem (see Eq. (23) in \cite{yurovsky15}). Then the imaginary part (\ref{ImSigma}), averaged over $r$, can be calculated with the sum rules (see Eq. (37) in  \cite{yurovsky15}), leading to
\bea
\Gamma_{S,S^z}=\frac{1}{f_S(N)}\sum_r \Gamma_{\mathcal{S}}=
\pi \sum_{S'}
\Bigl [ \delta_{S',S} Y^{(S,1)}[\hat{U}_{0},\hat{U}_{0}]
+\delta_{S',S+1} Y^{(S+1,1)}[\hat{U}_{-1},\hat{U}_{-1}]
+\delta_{S',S-1} Y^{(S,1)}[\hat{U}_{-1},\hat{U}_{-1}]\frac{f_{S-1}(N)}{f_S(N)}\Bigr]
\nonumber
\\
\times\left(X_{S_{z}0}^{(S,S',1)}\right)^2\sum_{\{v\},\{v'\}}\frac{1}{N}\sum_{i}|\langle v'_{i}|U|v_{i}\rangle|^{2}\prod_{i'\neq i}\delta_{v'_{i'},v_{i'}}
\delta(E_{S' ,S^z}-E_{S, S^z}+\sum_i(E_{v'_i}-E_{v_i})).
\phantom{xxx}
\eea
The width contains three terms, corresponding to $S'=S,S\pm1$, in agreement with the selection rules \cite{yurovsky14} for the case of one-body interactions.
Substituting the factors $X$ and $Y$ (see  Eq. (23) and Table I in \cite{yurovsky15}) and introducing the spectral function
\be
\label{specfunA}
\mathcal{A}(E) =\sum_{v,v'}n_B(E_v)|\langle v'|U|v\rangle|^{2}\delta(E+E_{v'}-E_v),
\ee
we get Eq. (\ref{result}).

\section{Level corrections}
The real part $\Sigma'$ of the self-energy $\Sigma = \Sigma' + i\Sigma''$ provides information on shifts of energy levels of the spin states due to their coupling to the vibrations.
We now demonstrate that these level shifts do not destroy the $\mathcal{O}(I N)$ energy separation of the large spin state from the remaining levels.

The summation over $\{v\},\{v'\},r'$ and averaging over $r$ is done in the same way as for the imaginary part in previous appendix. Then we get Eq. (\ref{result}), where the spectral function (\ref{specfunA}) is replaced by
\be
\label{sumReSigma}
\sum_{v,v'}n_B(E_v)\frac{|\langle v'|U|v\rangle|^{2}}{E_{S' ,S^z}-E_{S ,S^z}+E_{v'}-E_v} .
\ee

\begin{enumerate}
\item{Diagonal virtual transitions $S'=S$: in this case the denominator in Eq.~(\ref{sumReSigma}) is  the vibration energy difference and is independent of the spin state. The factor from Eq.~(\ref{result})  gives a correction to all energy levels
\be
\delta E \propto (\frac{N}{2}+1)\frac{(S^z)^2}{S(S+1)}.
\ee
This is negligible for unpolarized states $S^z \ll S$ by a factor of $1/N$.}

\item{Off-diagonal virtual transitions $S'=S \pm 1$: in this case the energy denominator is dominated by $E_{S\pm 1, S^z}-E_{S ,S^z} \sim I S$ leading to
\be
\delta E \propto (\frac{N}{2}+S+1)\frac{S^2-{S^z}^2}{(2S+1) S^2}
-(N/2-S)\frac{(S+1)^2-{S^z}^2}{(2S+1)S(S+1)} .
\ee
For the large-$S$ states the first term dominates and this correction is of $\mathcal{O}(1)$, much smaller than the energy difference between different $S$ states $ I S$. Furthermore, the dependence of this correction on $S^z$, is only of order $(S^z/S)^2$, again strongly suppressed for unpolarized states.}
\end{enumerate}

Thus the large spin state remains well separated from the smaller spin states: its separation from the next levels is of $\mathcal{O}(N)$, while the level corrections are of $\mathcal{O}(1)$.

\section{Emission rate}
We compute the transition rate between initial ($i$) and final ($f$) states where ${S^z}^{(i)} ={S^z}^{(f)}+1$. It is related to the matrix element $(i,f)$ of the evolution operator as $\Gamma_{i \to f} = \frac{d}{dt} |U_{fi}(t)|^2$. The transition amplitude occurs due to the perturbation $\delta H=\Omega e^{i E t} S^-+h.c.$, corresponding to coupling with the emitted electromagnetic radiation.

Denoting the free evolution operator of the decoupled spin system by $\mathcal{U}_0$, the rate of the transition in the absence of vibrations can be written as
\bea
\frac{d}{dt} |U_{fi}(t)|^2&=& \frac{d}{dt} \left[ \langle f | \int_{0}^{t} dt'  \mathcal{U}_0(t,t')V(t')\mathcal{U}_0(t',0) | i \rangle \langle i | \int_{0}^{t} dt''  \mathcal{U}_0(0,t'')V(t'')\mathcal{U}_0(t'',t) | f \rangle\right] \nonumber \\
&=& \langle f |  \int_{0}^{t} dt'  \mathcal{U}_0(t,t')V(t')\mathcal{U}_0(t',0) | i \rangle \langle i | \mathcal{U}_0(0,t)V(t)\mathcal{U}_0(t,t) | f \rangle + c.c.\nonumber \\
&=& \Omega^2 |\langle f |S^- | i \rangle|^2  e^{- iE t}   \int_{0}^{t} dt' \langle f| \mathcal{U}_0(t,t')|f\rangle e^{iE t'}\langle i|\mathcal{U}_0(t',t) | i \rangle + c.c.
\eea
The free evolution operator is related to the free resolvent
\begin{equation}\label{resolvap}
\mathcal{U}_0(t,0) = \frac{1}{2\pi i} \int_{-\infty}^\infty dz e^{-izt}R_0(z - i\eta),~~(t>0),
\end{equation}
where $R_0(z -i\eta) =(z - H_0 -i\eta)^{-1}$. We substitute this expression, perform the $t'$ integral, and use $\frac{e^{izt}-1}{iz}=\pi \delta(z)$ for large $t$. Noting that $\mathcal{U}_0(t',t)$ corresponds to evolution backwards in time and corresponds to the complex conjugate of Eq.~(\ref{resolvap}), we have
\be
\frac{d}{dt} |U_{fi}(t)|^2=-\frac{1}{2i}\Omega^2 |\langle f |S^- | i \rangle|^2  \int_{-\infty}^\infty \frac{dz}{2\pi i} \langle f | R(z-i\eta)|f\rangle \langle i|R(z + E + i\eta) | i \rangle + c.c.
\ee
We now incorporate the influence of vibrations on this result by inserting the vibration-induced self-energy in the resolvents [Eq.~(\ref{Resolvent})],
\bea
\frac{d}{dt} |U_{if}(t)|^2 &=&-\frac{1}{2i}\Omega^2 |\langle f |S^- | i \rangle|^2  \int_{-\infty}^\infty \frac{dz}{2\pi i} \frac{1}{z - E_f - \Sigma_f} \frac{1}{z -E- E_i - \Sigma_i^*} + c.c. \nonumber \\
&=& -\frac{1}{2i}\Omega^2 |\langle f |S^- | i \rangle|^2  \frac{1}{E_i + \Sigma_{i}' - E_f - \Sigma_f' -E+ i (\Gamma_i+\Gamma_f)} + c.c. \nonumber \\
&=& \Omega^2 |\langle f |S^- | i \rangle|^2 \frac{\Gamma_i+\Gamma_f}{(E_i + \Sigma_{i}' - E_f - \Sigma_f'+E)^2 + (\Gamma_i+\Gamma_f)^2},
\eea
where $\Sigma_{i,f} = \Sigma_{i,f}' + i\Gamma_{i,f}$ is the self-energy. Thus we obtain Eq.~(\ref{gammaif}).

\end{widetext}

\end{document}